\documentclass[aps,twocolumn,superscriptaddress,groupedaddress,nofootinbib]{revtex4}  

\usepackage{graphicx}  
\usepackage{bm}     
\usepackage{amssymb}  
\usepackage[utf8]{inputenc}
\usepackage{amsmath}
\usepackage{xspace}
\usepackage{physics}

\usepackage{hyphenat}
\usepackage{hyperref}
\hypersetup{
    colorlinks=true,
    allcolors=[rgb]{0.26,0.41,0.88},
}

\usepackage{newtx}

\newcommand{\TS}{\texttt{\allowbreak Trans\-ition\-Sol\-ver}}
\newcommand{\PT}{\texttt{\allowbreak Phase\-Tra\-cer}}
\newcommand{\CT}{\texttt{\allowbreak Cosmo\-Trans\-itions}}

\newcommand{\vecb}[1]{\boldsymbol{#1}}
\newcommand{\unit}[1]{\unskip\ensuremath{\,\text{#1}}}
\newcommand{\tentothe}[1]{\!\times\!10^{#1}}

\newcommand{\rhotot}{\rho_{\text{tot}}}
\newcommand{\rhogs}{\rho_{\text{gs}}}
\newcommand{\field}{\vecb{\phi}}
\newcommand{\Vext}{\mathcal{V}_t^{\text{ext}}}
\newcommand{\pseudotrace}{\bar{\theta}}
\newcommand{\tausw}{\tau_{\text{sw}}}
\newcommand{\overbar}[1]{\mkern 3mu\overline{\mkern-3mu#1\mkern-1.5mu}\mkern 1.5mu}
\newcommand{\Uf}{\overbar{U}_{\!f}}
\newcommand{\Treh}{T_{\text{reh}}}

\newcommand{\mev}{\,\text{MeV}\xspace}
\newcommand{\gev}{\,\text{GeV}\xspace}
\newcommand{\nhz}{\,\text{nHz}\xspace}

\usepackage{cleveref}

\newcommand{\smref}[1]{\cref{#1}}

\begin{document}
\title{Can supercooled phase transitions explain the gravitational wave background observed by pulsar timing arrays?}

\author{Peter Athron}
\email[]{peter.athron@njnu.edu.cn}
\affiliation{Department of Physics and Institute of Theoretical Physics, Nanjing Normal University, Nanjing, 210023, China}

\author{Andrew Fowlie}
\email[]{andrew.fowlie@xjtlu.edu.cn}
\affiliation{Department of Physics, School of Mathematics and Physics, Xi'an Jiaotong-Liverpool University, Suzhou, 215123, China}

\author{Chih-Ting Lu}
\email[]{ctlu@njnu.edu.cn}
\affiliation{Department of Physics and Institute of Theoretical Physics, Nanjing Normal University, Nanjing, 210023, China}

\author{Lachlan Morris}
\email[]{lachlan.morris@monash.edu}
\affiliation{School of Physics and Astronomy, Monash University, Melbourne, Victoria 3800, Australia}

\author{Lei Wu}
\email[]{leiwu@njnu.edu.cn}
\affiliation{Department of Physics and Institute of Theoretical Physics, Nanjing Normal University, Nanjing, 210023, China}

\author{Yongcheng Wu}
\email[]{ycwu@njnu.edu.cn}
\affiliation{Department of Physics and Institute of Theoretical Physics, Nanjing Normal University, Nanjing, 210023, China}

\author{Zhongxiu Xu}
\email[]{zhongxiuxu@njnu.edu.cn}
\affiliation{Department of Physics and Institute of Theoretical Physics, Nanjing Normal University, Nanjing, 210023, China}

\begin{abstract}

Several pulsar timing array collaborations recently reported evidence of a stochastic gravitational wave background (SGWB) at nHz frequencies.
Whilst the SGWB could originate from the merger of supermassive black holes, it could be a signature of new physics near the 100\mev scale.
Supercooled first-order phase transitions (FOPTs) that end at the 100\mev scale are intriguing explanations, because they could connect the nHz signal to new physics at the electroweak scale or beyond. 
Here, however, we provide a clear demonstration that it is not simple to create a nHz signal from a supercooled phase transition, due to two crucial issues that 
could rule out many proposed supercooled explanations and should be checked.
As an example, we use a model based on non-linearly realized electroweak symmetry that has been cited as evidence for a supercooled explanation.
First, we show that a FOPT cannot complete for the required transition temperature of around 100\mev.
Such supercooling implies a period of vacuum domination that hinders bubble percolation and transition completion.
Second, we show that even if completion is not required or if this constraint is evaded, the Universe typically reheats to the scale of any physics driving the FOPT.
The hierarchy between the transition and reheating temperature makes it challenging to compute the spectrum of the SGWB.
\end{abstract}

\maketitle
\section{Introduction}

NANOGrav recently detected a stochastic gravitational wave background (SGWB) for the first time with a significance of about $4\sigma$~\cite{NANOGrav:2023gor}. This was corroborated by other pulsar timing arrays (PTAs), including the CPTA~\cite{Xu:2023wog}, EPTA~\cite{Antoniadis:2023ott} and PPTA~\cite{Reardon:2023gzh}. Although the background could originate from mergers of super-massive black holes~\cite{NANOGrav:2023hfp,Ellis:2023dgf}, this explanation might be inconsistent with previous estimates of merger
density and remains a topic of debate~\cite{Casey-Clyde:2021xro,Kelley:2016gse,Kelley:2017lek,Shen:2023pan}. Thus, there is an intriguing possibility that the SGWB detected by NANOGrav could originate from more exotic sources~\cite{NANOGrav:2023hvm}. Indeed, many exotic explanations were proposed for an earlier hint of this signal~\cite{NANOGrav:2021flc,NANOGrav:2020bcs,Bian:2020urb}, or immediately after the announcement. These include non-canonical kinetic terms~\cite{Yi:2021lxc}, inflation~\cite{Vagnozzi:2020gtf,Benetti:2021uea,Gao:2021lno,Ashoorioon:2022raz,Vagnozzi:2023lwo}, first-order phase transitions~(FOPTs; \cite{Nakai:2020oit,Ratzinger:2020koh,Xue:2021gyq,Deng:2023seh,Megias:2023kiy}), cosmic strings~\cite{Blasi:2020mfx,Ellis:2020ena,Buchmuller:2020lbh,Blanco-Pillado:2021ygr,Bian:2022tju,Samanta:2020cdk,Wang:2023len,Ellis:2023tsl}, domain walls \cite{Ferreira:2022zzo,King:2023cgv},  primordial black holes~\cite{Franciolini:2023pbf}, primordial magnetic fields \cite{Li:2023yaj}, axions and ALPs~\cite{Ramberg:2020oct,Inomata:2020xad,Sakharov:2021dim,Kawasaki:2021ycf,Guo:2023hyp,Kitajima:2023cek,Yang:2023aak}, QCD~\cite{Neronov:2020qrl,Bai:2023cqj}, and dark sector models~\cite{Addazi:2020zcj,Li:2021qer,Borah:2021ocu,Borah:2021ftr,Freese:2022qrl,Freese:2023fcr,Fujikura:2023lkn,Zu:2023olm,Han:2023olf}.

The nanohertz (nHz) frequency of the signal indicates that any new physics explanation should naturally lie at around 100\mev. If there are new particles around the MeV scale there are constraints from cosmology~\cite{Foot:2014uba,Bai:2021ibt,Bringmann:2023opz,Madge:2023cak} and, in any case, from particle physics experiments. It is conceivable, however, that new physics at characteristic scales far beyond the MeV scale could be responsible for a nHz signal. This could happen, for example, if a FOPT \cite{Caprini:2015zlo, Caprini:2019egz, Athron:2023xlk} starts at higher temperatures but supercools down to $100\mev$. 
That is, the Universe remains in a false vacuum until the $100\mev$ scale because a transition to the true vacuum is suppressed.

This was previously considered for an electroweak phase transition \cite{Kobakhidze:2017mru,Witten:1980ez, Iso:2017uuu, Arunasalam:2017ajm, vonHarling:2017yew, Baratella:2018pxi, Bodeker:2021mcj, Sagunski:2023ynd} and was discussed as a possible new physics explanation by NANOGrav~\cite{NANOGrav:2023hvm,NANOGrav:2021flc}. Supercooling could help new physics explanations evade constraints on MeV-scale modifications to the SM and connect a nHz signal to new physics and phenomenology at the electroweak scale or above. 

In this {\it Letter}, however, we raise two difficulties with supercooled FOPTs. We explicitly demonstrate that these difficulties rule out one of the prominent models that explain the nHz GW signal through a supercooled FOPT used in Refs.~\cite{NANOGrav:2023hvm,NANOGrav:2021flc}. Firstly, the phase transition does not complete for the low temperatures associated with a nHz signal. This finding is consistent with brief remarks in Ref.~\cite{Ellis:2018mja} and, as mentioned there, similar to the graceful exit problem in old inflation~\cite{Guth:1982pn}. Secondly, the energy released by the phase transition reheats the Universe to about the new physics scale~\cite{Madge:2023cak} and this can rule out attempts to solve the completion problem. However we also show that for supercooled phase transitions the temperature dependence is more complicated than naive arguments suggest, and the hierarchy between the percolation and reheating temperatures must be taken into account when computing the SGWB spectrum.

\section{Cubic potential and Benchmarks}
\label{sec:supercool_model}
We consider a modification to the SM Higgs potential to include a cubic term,
\begin{equation}
    V_0 (r) = -\frac{\mu^2}{2} r^2 +\frac{\kappa}{3}r^3 +\frac{\lambda}{4}r^4.
\end{equation}
For further details about the model and effective potential, see \smref{app:potential_rad_correc} and Ref.~\cite{Kobakhidze:2016mch,Kobakhidze:2017mru,Parwani:1991gq,Ekstedt:2022bff,Coleman:1973jx}. We define the percolation temperature, $T_p$, and completion temperature, $T_f$, of a transition as the temperatures at which the false vacuum fraction $P_f = 0.71$ and $0.01$, respectively~(\cite{doi:10.1063/1.1338506, LIN2018299, LI2020112815}; see \smref{app:phaseTransitionAnalysis} for further details about phase transition analysis, which includes Refs.~\cite{Athron:2020sbe,Athron:2022jyi,Athron:2023ts,Wainwright:2011kj,Athron:2022mmm,Athron:2023xlk,Linde:1981zj,Wu:2019pbm,Espinosa:2010hh,Turner:1992tz}).

We consider two benchmark points to highlight the challenges of fitting a nHz signal with this cubic potential. These benchmarks are selected to probe two criteria: 1) realistic percolation, that is, having a percolation temperature and that the physical volume of the false vacuum is decreasing at the onset of percolation; and 2) having a completion temperature, that is, a temperature at which the false vacuum fraction falls to below $1\%$. These benchmarks are:
\begin{equation}
    \text{BP1: } \kappa = -117.96\gev, \quad 
    \text{BP2: } \kappa = -118.67\gev. 
\end{equation}
BP1 resulted in the most supercooling for which the transition satisfies both criteria, though fails to supercool to sub-GeV temperatures. For BP1, the physical volume of the false vacuum starts decreasing at exactly the percolation temperature. Increasing supercooling any further thus violates our first criteria. BP2 resulted in stronger supercooling with a nominal percolation temperature of 100\mev but no completion temperature. However, although BP2 was chosen so that percolation was estimated to begin at 100\mev, it violates our first criteria and the space between bubbles continues to expand below 100\mev. Thus, despite a nominal percolation temperature, percolation could be unrealistic. Without significant percolation of bubbles, the phase transition would not generate a SGWB. The benchmarks are sensitive to uncertainties; for example, changing the Higgs mass by $1\sigma$, $0.17\gev$~\cite{Workman:2022ynf}, changes the value of $\kappa$ below which percolation is unrealistic (BP1) and the associated percolation temperature by about $0.5\gev$. Our conclusions and results, however, would be qualitatively unchanged.

\section{Challenges}
\subsection{Challenge 1 --- percolation and completion}\label{sec:completion}

As discussed, supercooling was proposed to achieve a peak frequency at the nHz scale. However, in many models, a first-order electroweak phase transition has bubbles nucleating at around the electroweak scale. There is then an extended period of bubble growth and expansion of space. If bubbles grow too quickly compared to the expansion rate of the Universe, the bubbles will percolate before sufficient supercooling. Yet if bubbles grow too slowly the transition may never percolate or complete due to the space between bubbles inflating~\cite{Turner:1992tz, Ellis:2018mja, Athron:2022mmm}; this effect can cause both the realistic percolation condition and the condition for a completion temperature to fail. Thus, while it is possible to tune model parameters to achieve a nominal percolation temperature at sub-GeV temperatures, true percolation and completion of the transition become less likely as supercooling is increased. 

We find that a completion temperature is impossible for the cubic potential if $T_p \lesssim 1\gev$. The same arguments apply to the models considered in Ref.~\cite{Athron:2022mmm}. In the cubic potential, strong supercooling implies a Gaussian bubble nucleation rate peaking at $T_\Gamma \sim 50\gev$.\footnote{It might be possible to evade this argument in models that predict a non-Gaussian nucleation rate, e.g., conformal models~\cite{Jinno:2016knw, Marzo:2018nov, Ellis:2019oqb, Ellis:2020nnr}.} 

In Ref.~\cite{NANOGrav:2023hvm}, the cubic potential is suggested as a candidate model for a strongly supercooled phase transition that could explain the detected SGWB. The Universe was assumed to be radiation dominated in the original investigation \cite{Kobakhidze:2017mru} of detecting GWs from the cubic potential with PTAs. However, a more careful treatment of the energy density during strong supercooling shows that the Universe becomes vacuum dominated \cite{Ellis:2018mja}. This leads to a prolonged period of rapid expansion that hinders bubble percolation and completion of the transition. In fact, one must check not only that $P_f < 0.01$ eventually, but also that the physical volume of the false vacuum is decreasing at $T_p$~\cite{Turner:1992tz, Ellis:2018mja, Athron:2022mmm}.

\begin{figure}
    \centering
    \includegraphics[width=0.9\linewidth]{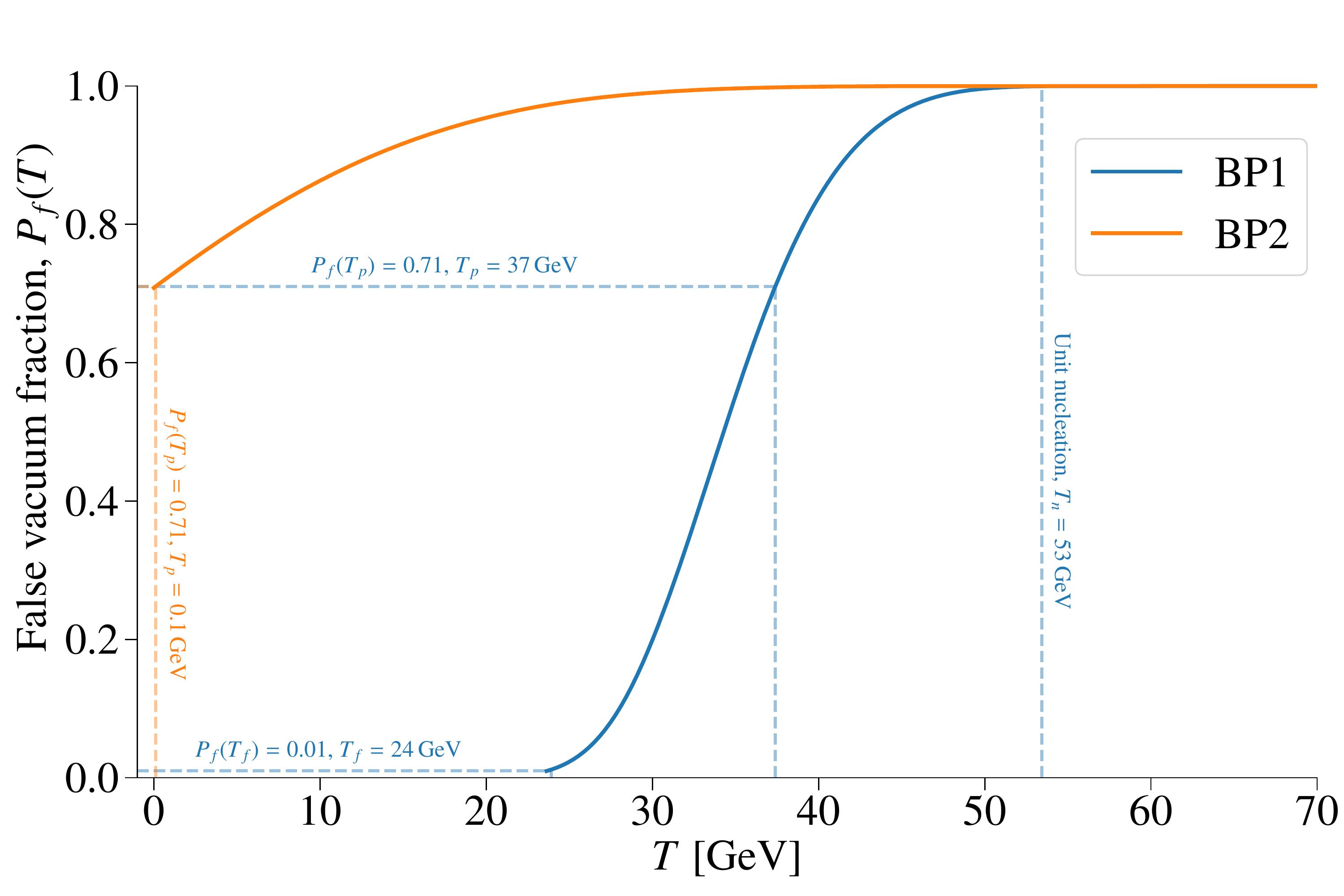}
    \caption{The false vacuum fraction as a function of temperature for BP1 (blue, right-most solid curve) and BP2 (orange, left-most solid curve). The nucleation, percolation and completion temperatures are shown for BP1. However, BP2 only has a percolation temperature at $T_p \approx 100\mev$.}
    \label{fig:Pf_vs_T}
\end{figure}

The SGWB from a FOPT should not be computed at the nucleation temperature $T_n$, as this will generally give a very different result compared to computing it at lower temperatures where bubbles are actually colliding~\cite{Athron:2023rfq}. The percolation temperature is a much better choice~\cite{Athron:2022mmm}. By definition, we anticipate the formation of a cluster of connected bubbles at the percolation temperature and thus bubble collisions and the generation of GWs are expected to begin at around this time. Fig.\ \ref{fig:Pf_vs_T} demonstrates the large difference between $T_n$ and $T_p$ in supercooled phase transitions. In BP1 the difference is $\mathcal{O}(10\gev)$. In BP2 there is no nucleation temperature --- one might be tempted to assume GWs cannot be produced because of this. However, percolation and completion are possible even without a nucleation temperature \cite{Athron:2022mmm}. Another large source of error is the use of $\beta/H$ for estimating the timescale of the transition. The mean bubble separation can be used instead as described in \smref{app:thermalParameters} for thermal parameters, which includes Refs.~\cite{Athron:2022mmm,Giese:2020rtr,Laurent:2022jrs,Athron:2023xlk,Giese:2020znk,Cutting:2019zws,Megevand:2016lpr,Ellis:2018mja,Guo:2020grp}.

\subsection{Challenge 2 --- reheating}\label{sec:reheating}

Even if the completion constraints can be avoided, a second issue was recently observed~\cite{Madge:2023cak}.  Whilst strong supercooling can lower the percolation temperature down to $T_p \approx 100\mev$ as in BP2 or even lower, the energy released from the phase transition reheats the plasma, creating a hierarchy $\Treh \gg T_p$.  Indeed, the reheating and percolation temperatures are approximately related by~\cite{Ellis:2018mja}
\begin{equation}\label{eq:approx}
    \Treh \simeq (1 + \alpha(T_p))^{1/4} \, T_p,
\end{equation}
where $\alpha$ is the transition strength and is related to the latent heat released during the phase transition. The substantial latent heat in a strongly supercooled transition, $\alpha \gg 1$, thus implies that $\Treh \gg T_p$.  Ref.~\cite{Madge:2023cak} approximate $\alpha \approx \Delta V / \rho_R$ from the free energy difference ($\Delta V$) and the radiation energy density ($\rho_R$) and shows that in the Coleman-Weinberg model the latent heat is so large that the Universe reheats well above the percolation temperature and back to the scale of new physics.

A simple scaling argument suggests that this observation --- that supercooled FOPTs reheat to the scale of new physics, $M$ --- is generic. The new physics creates a barrier between minima so we expect $\Delta V \sim M^4$, and because the radiation energy density goes like $T_p^4$, we expect the latent heat may go like $\alpha \sim {M^4}/{T_p^4}$. This leads to
 \begin{equation}\label{eq:scaling_argument}
     \Treh \sim \left(\frac{M^4}{T_p^4}\right)^{\frac14} T_p = M .
 \end{equation}
It is possible that reheating to $\Treh \ll M$ could be achieved, however, by avoiding $\Delta V \sim M^4$. For example, by fine-tuning couplings in the potential such that, despite new physics at a scale $M$ creating a second minima separated by a barrier, the relative depth of the minima at $T_p \ll M$ is much less than $M^4$ such that $\Delta V \ll M^4$.

The arguments leading to \cref{eq:scaling_argument}, however, rely on the simple approximation of the reheating temperature in~\cref{eq:approx} and crude dimensional analysis. We now confirm that this problem exists and is unavoidable in a careful analysis of the example model we consider. This careful treatment is general and can be used in other models.  We assume that the reheating occurs instantaneously around the time of bubble percolation, and use conservation of energy so that the reheating temperature can be obtained from~\cite{Athron:2023ts,Athron:2022mmm}
\begin{equation}\label{eq:reheat}
\rho(\phi_f(T_p), T_p) = \rho(\phi_t(\Treh ), \Treh ) ,
\end{equation}
where $\phi_f$ and $\phi_t$ are the false and true vacua, respectively, and $\rho$ is the energy density. For BP1, the percolation temperature is $T_p \approx 37.4\gev$,  and the transition completes and reheats the Universe to $\Treh \approx 44.1\gev$. The reheating temperature exceeds the percolation temperature due to the energy released from the supercooled phase transition, though they remain of the same order of magnitude.
For BP2, however, the percolation temperature drops to $T_p \approx 100\mev$, whereas the reheating temperature is $\Treh \approx 35.6\gev$; more than two orders of magnitude larger.

\begin{figure}
   \centering
    \includegraphics[width=0.9\linewidth]{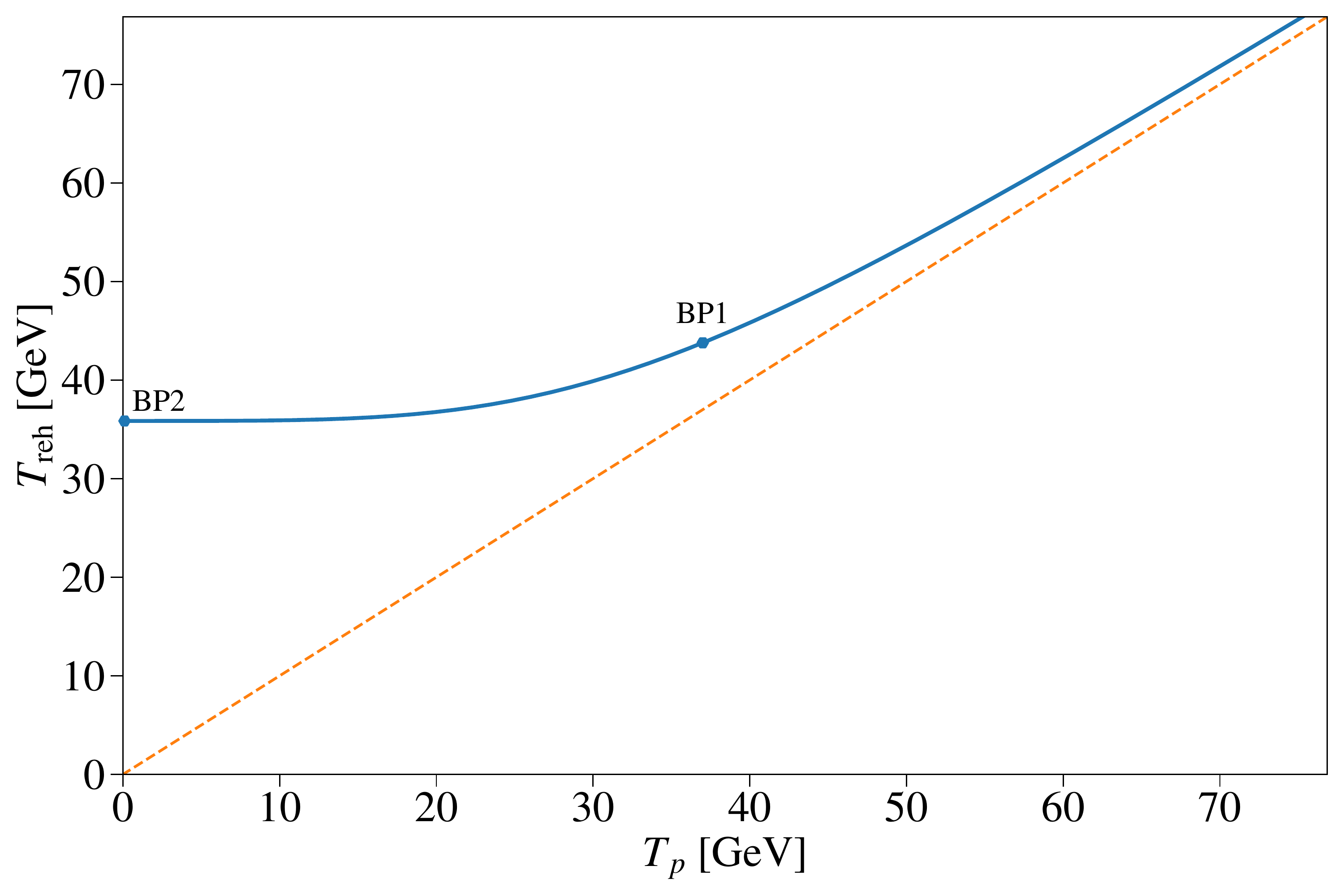}
    \caption{The reheating temperature $\Treh$ against percolation temperature $T_p$ as $\kappa$ varied. The dashed orange line corresponds to $\Treh = T_p$. We see that $\Treh \gtrsim 36\gev$ even when $T_p \to 0$. Our two benchmark points are labeled.}
    \label{fig:TrehVSTp}
\end{figure}

We show the behavior of the reheating temperature as a function of percolation temperature in \cref{fig:TrehVSTp}. We clearly see that the reheating temperature tends towards a constant value $\Treh \approx 36\gev$ for $T_p \to 0$. As we now discuss, the fact that $T_\Gamma \sim \Treh  \gg T_p$ breaks assumptions typically made when computing the SGWB.

\subsection{Gravitational Wave Spectra}
\label{sec:GWsResults}

The frequencies of a SGWB created at a percolation temperature $T_p$ would be redshifted from the reheating temperature $\Treh$ to the current temperature $T \simeq 2.725 \unit{K}$ \cite{Athron:2023xlk}. The redshifted peak frequency of the SGWB today would be
\begin{equation}
    f_p \approx
    10 \unit{nHz} \,\left(\vphantom{\frac{f_\text{peak}|_{T=T_p}}{H_*}}\frac{g_*(\Treh)}{100}\right)^{\!\frac16} \left(\vphantom{\frac{f_\text{peak}|_{T=T_p}}{H_*}}\frac{\Treh}{100 \unit{MeV}} \right) \left(\frac{1}{R_* H(\Treh)} \right) , \label{eq:freqRedshift-new}
\end{equation}
where $R_*$ is the mean bubble separation, $H$ is the Hubble parameter and $g_*$ is the number of effective degrees of freedom.\footnote{We apply suppression factors from Ref.~\cite{Husdal:2016haj} to the degrees of freedom of each particle when estimating $g_*$. This incorporates the effects of particles decoupling from the thermal bath as the temperature drops below their respective mass. The peak frequency and amplitude depend only weakly on $g_*$ such that mismodeling $g_*$ cannot dramatically change the SGWB.}
In the absence of supercooling we anticipate that $\Treh \sim T_p$, such that $R_* H(\Treh) \sim R_* H(T_p)$ and since the bubbles would not have a long time to grow $R_* H(T_p) \lesssim 1$.  Thus, in the absence of supercooling, we expect $T_p \sim 100 \mev$ to lead to a $\sim\!\!10\nhz$ signal.

In this cubic model, however, $T_p \sim 100\mev$ requires strong supercooling, so we now consider an analysis more appropriate for this scenario. At the time of the phase transition the peak frequency $f_{p,*}$ is set by the mean bubble separation $R_*$ via $f_{p,*}\sim 1/R_*$ \cite{Caprini:2015zlo}. After redshifting, the peak frequency of the SGWB today scales as
\begin{equation}\label{eq:f_scale_supercool}
    f_p \sim \frac{1 \unit{GeV}} {R_*(T_p) s_t(\Treh)^{1/3}},    
\end{equation}
where $s_t$ is the true vacuum entropy density (see \smref{app:redshift}). Because radiation domination is a valid assumption in the true vacuum, the entropy density scales as $s_t(T) \sim T^3$.

One can show that $R_* \sim 1 \unit{GeV} / (T_\Gamma T_p N^{1/3})$ if bubbles nucleate simultaneously at $T_\Gamma$, 
where $N$ is the total number of bubbles nucleated per Hubble volume throughout the transition.\footnote{Simultaneous nucleation is an extreme case of Gaussian nucleation, found to be a good approximation in this model~\cite{Athron:2022mmm}.}Combining this with \cref{eq:f_scale_supercool}, we obtain
\begin{equation}\label{eq:freq_scaling_argument}
    f_p \sim T_p \left(\frac{T_\Gamma N^\frac13}{\Treh}\right).
\end{equation}
Numerically, we find that $N^\frac13$, $T_\Gamma$ and  $\Treh$ -- and thus the right-most factor --- depend only weakly on the amount of supercooling (see \cref{fig:TrehVSTp}). Thus, for supercooling we find the relationship $f_p \sim T_p$. This suggests that one can obtain an arbitrarily low peak frequency by fine-tuning the percolation temperature.

In the cubic model, these arguments are surprisingly accurate. Indeed, we find numerically that
\begin{equation}\label{eq:radius_approx}
\frac{1}{R_* H(\Treh)} \simeq 1.1 \, \left(\frac{T_p}{\Treh}\right) \left(\frac{T_\Gamma N^\frac13}{\Treh}\right).
\end{equation}
Assuming radiation domination in the true vacuum for $H(\Treh)$ and that $g_* \approx 100$, \cref{eq:radius_approx,eq:freqRedshift-new} lead to 
\begin{equation}\label{eq:f_peak_supercooling}
   f_p \approx
   10 \unit{nHz} \, \left(\vphantom{\frac{N^\frac13}{\Treh^2}} \frac{T_p}{100 \unit{MeV}} \right) \left(\frac{T_\Gamma N^\frac13}{\Treh} \right) ,
\end{equation}
in agreement with the scaling anticipated in \cref{eq:freq_scaling_argument}. The right-most factor in \cref{eq:radius_approx,eq:f_peak_supercooling} is $\mathcal{O}(1)$ and approximately independent of the amount of supercooling. Thus, to achieve a redshifted peak frequency of $10\nhz$, we require $T_p \approx 100\mev$. 

Comparing \cref{eq:f_peak_supercooling} with the result in the absence of supercooling \cref{eq:freqRedshift-new}, supercooling and subsequent substantial reheating redshift the frequency more than usual. However, assuming radiation domination \cref{eq:radius_approx} leads to
\begin{equation}\label{eq:crude}
R_* H(T_p) \approx \frac{T_\Gamma}{T_p}.
\end{equation}
This increase in bubble radius caused by the delay between nucleation and percolation partially offsets the impact of additional redshifting.

Our findings are contrary to the claim in Refs.~\cite{Ellis:2018mja,Madge:2023cak} that reheating makes it difficult to reach GW frequencies relevant for PTAs. However, we do agree with the finding in Ref.~\cite{Ellis:2018mja} that completion poses an issue for nHz GW signals in this model. As found in \cref{sec:completion}, a percolation temperature of $T_p = 100\mev$ would not result in a successful transition. Not only would the majority of the Universe remain in the false vacuum even today, the true vacuum bubbles would not actually percolate due to the inflating space between the bubbles.

We now consider the SGWB predictions. We use the pseudotrace \cite{Giese:2020znk} to avoid assumptions about the speed of sound and the equation of state that can break down in realistic models. We also use the mean bubble separation rather than proxy timescales derived from the bounce action that are invalid for strongly supercooled phase transitions. For a full description, see \smref{app:thermalParameters,app:GWsFits} which includes Refs.~\cite{Cai:2017tmh,Athron:2023xlk,Fixsen:2009ug,ParticleDataGroup:2020ssz,Caprini:2015zlo,Caprini:2019egz,Lewicki:2022pdb,Hindmarsh:2016lnk,Hindmarsh:2019phv,Gowling:2022pzb,Hindmarsh:2017gnf,Cai:2023guc,Ghosh:2023aum,Caprini:2010xv,Caprini:2009yp}.

In this model we find that the bubbles mostly nucleate at temperatures around $T_\Gamma \sim 50 \gev$. We thus expect that friction from the plasma is sufficient to prevent runaway bubble walls, despite the large pressure difference.  This implies that the SGWB from bubble collisions is negligible and that all the available energy goes into the fluid, resulting in a SGWB from sound waves and turbulence.

\begin{figure}
    \centering
    
    \hbox{\put(105,0){\scriptsize{BP1 --- $T_p \simeq 37.4\gev$}}}
    \includegraphics[width=0.9\linewidth]{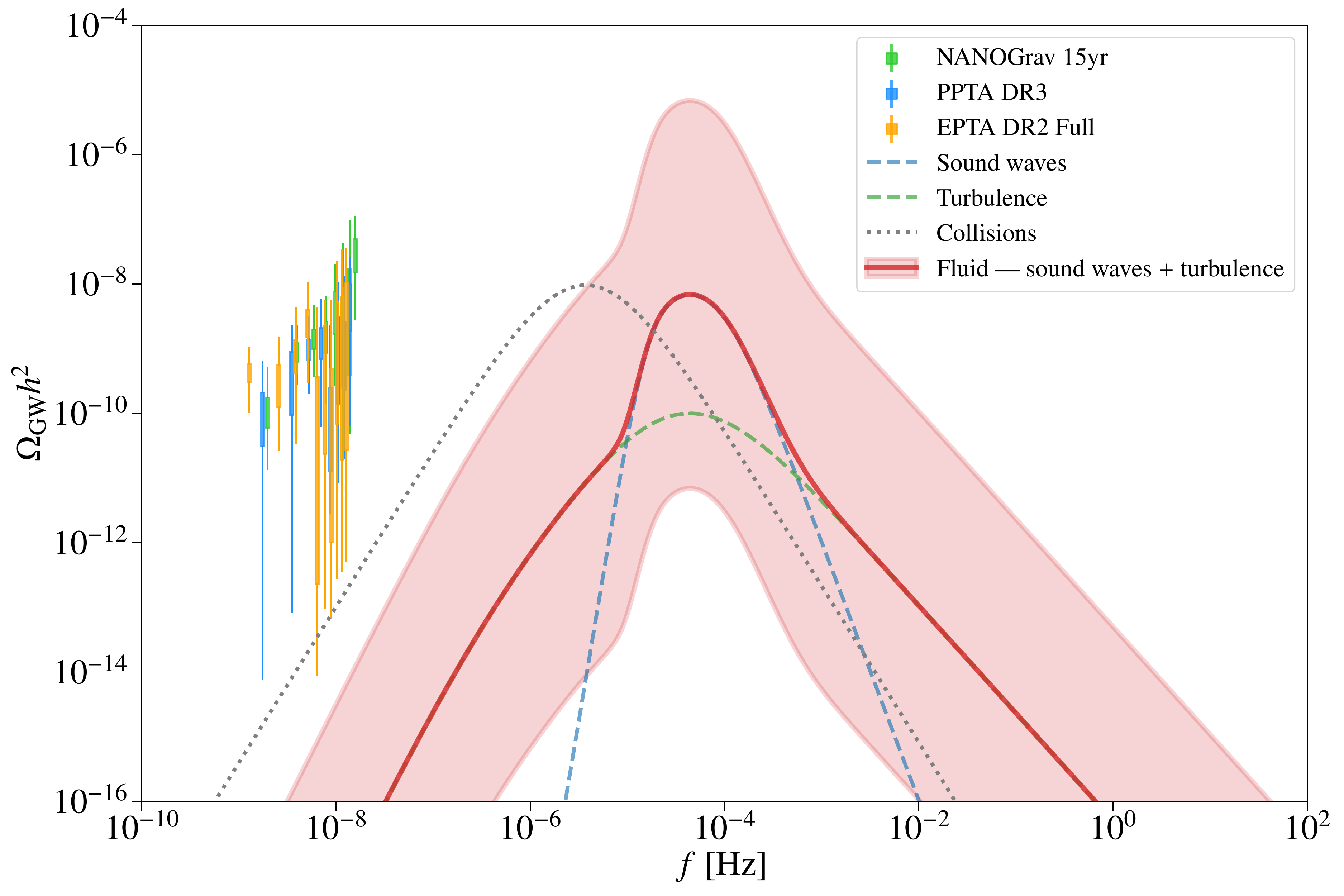}
    
    \vspace{0.25cm}

    \hbox{\put(85,0){\scriptsize{\textbf{Unphysical}. BP2 --- $T_p \simeq 100\mev$}}} 
    \includegraphics[width=0.9\linewidth]{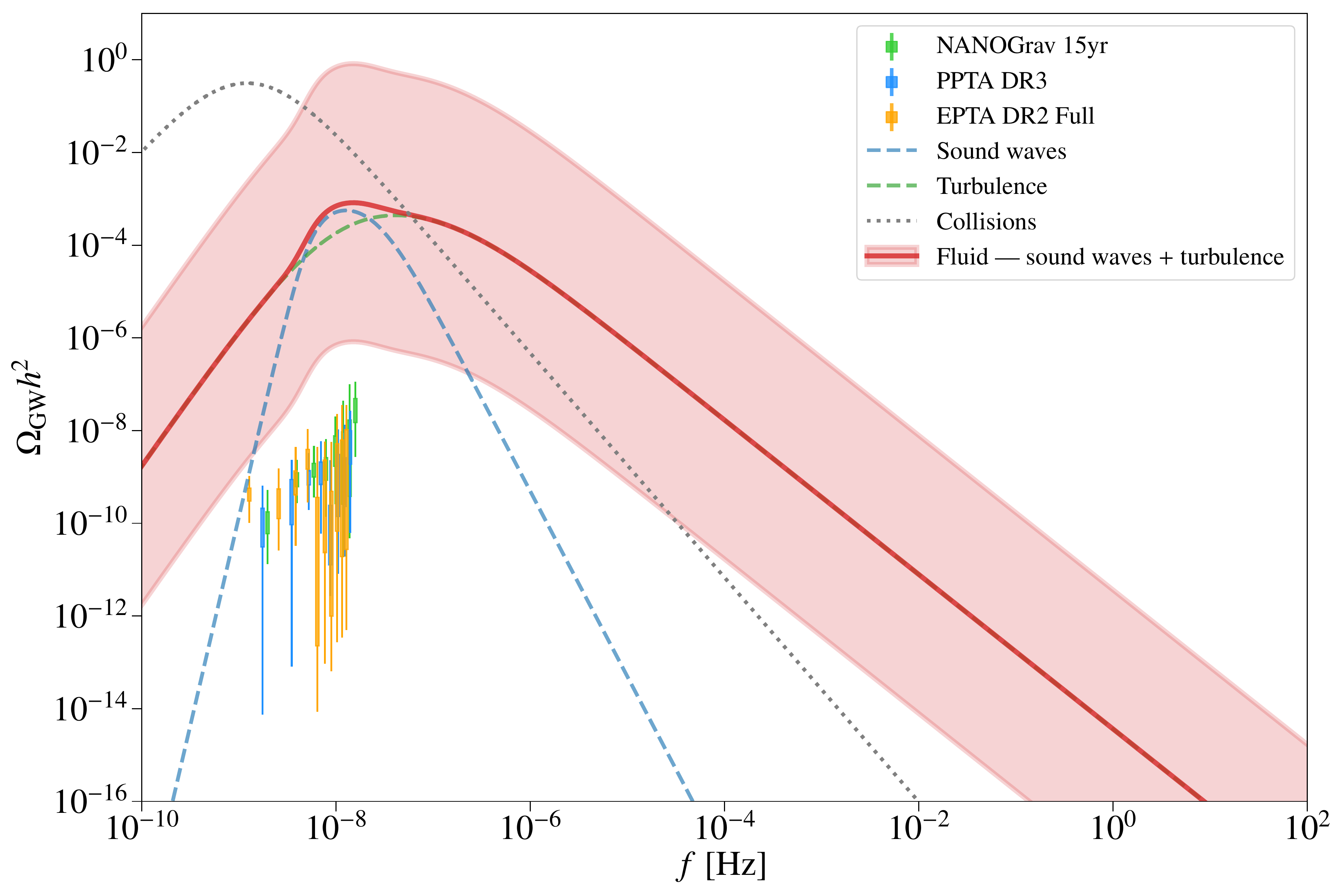}
    \caption{%
    The SGWB from BP1 (top panel; strongest supercooling for which the FOPT completed) and the unphysical SGWB from BP2 (bottom panel; strongest supercooling for which the FOPT has a percolation temperature though it does not complete and percolation is questionable).
    We show the $50\%$ and $95\%$ bands for the PTA observations (box plots).
    The BP1 and BP2 predictions fail to match the PTA observations, even when allowing for a factor $\mathcal{O}(10^3)$ uncertainty (shaded red band). 
    For our BPs, the total SGWB (solid red) comes only from sound waves (dashed blue) and turbulence (dashed green). 
    For comparison, we show the SGWB from a vacuum transition where bubble collisions would be the only source of GWs (dotted grey).}
    \label{fig:gws_BP1}
\end{figure}

In \cref{fig:gws_BP1} we show the predicted SGWB spectrum for both BP1 (upper panel; the model with maximal supercooling while guaranteeing percolation and completion) and  BP2 (lower panel; the model with a percolation temperature at 100\mev but questionable percolation and no completion). The peak frequencies are about $4\!\times\! 10^{4}\nhz$ and $15\nhz$ for BP1 and BP2, respectively. BP1 represents the lowest peak frequency that can be obtained for realistic scenarios in this model because for more supercooling the transition does not complete and percolation becomes questionable. To compare the BP1 predictions with the PTA signals, we must consider the theoretical uncertainties.  In our analysis we used daisy resummation and full one-loop corrections to the effective potential. Whilst this approach suffers from substantial theoretical uncertainties, leading to a factor $\mathcal{O}(10^3)$ uncertainty in the predicted GW amplitude~\cite{Croon:2020cgk}, the BP1 predictions lie more than 7 orders of magnitude below the NANOGrav signal at nHz frequencies. Thus this model \emph{cannot} explain the nHz signal observed by PTA experiments despite various optimistic statements from the literature. For comparison we show the SGWB prediction if one were to assume vacuum transitions (dotted grey curves). This assumption is not realistic for this model and in any case does not result in agreement with the observed spectrum. 

If one ignores the percolation and completion requirements, BP2 shows that the peak frequency can be reduced to match the nHz signal observed by PTA experiments, though the amplitude is several orders of magnitude higher than the PTA observations. Caution should be taken interpreting the SGWB predictions for such strong supercooling because it is well beyond what has been probed in simulations. These predictions are somewhat unphysical because, despite a nominal percolation temperature, bubbles are not expected to percolate as the false vacuum between them is inflating. Without percolation, GWs would not be generated. Lastly, we note that points between BP1 and BP2 may exist in which the low-frequency tail of the SGWB passes through the PTA observations. The transitions for such points, however, would not complete.

\section{Conclusions}\label{sec:conclusions}

Supercooled FOPTs are an intriguing explanation of the nHz SGWB recently observed by several PTAs, as they could connect a nHz signal to the electroweak scale. Indeed, they were mentioned as a possibility~\cite{NANOGrav:2023gor,NANOGrav:2023hvm}. However we demonstrate two major difficulties that can affect supercooled explanations. First, percolation and completion of the transition are hindered by vacuum domination.  We demonstrate with an explicit numerical calculation that this rules out the possibility of explaining the PTA signal in the supercooling model of Ref.\ \cite{Kobakhidze:2017mru} mentioned as a prototypical example in Refs.~\cite{NANOGrav:2023gor,NANOGrav:2023hvm}. 

Second, the Universe typically reheats to the scale of any physics driving the transition, splitting the percolation and reheating temperatures significantly. This makes it challenging to compute the signal from a supercooled transition because factors often implicitly neglected must be carefully included in fit formulae and the thermal parameters are well beyond those in hydrodynamical simulations on which fit formulae are based.  The correct scaling, \cref{eq:freq_scaling_argument,eq:f_peak_supercooling}, shows that for supercooled phase transitions that do not complete, the peak frequency could be reduced to nHz. In contrast, completing the phase transition by increasing the nucleation rate at late stages would {\it not} lead to a nHz signal due to a higher bubble number density, ruling out solutions similar to those proposed for the graceful exit problem~\cite{Adams:1990ds,DiMarco:2005zn}.  We anticipate that these issues are quite generic and they should be carefully checked in supercooled explanations.

\begin{acknowledgments}
P.A., C.L. and L.W. are supported by the National Natural Science Foundation of China (NNSFC) under Grant No. 12335005. P.A. is also supported by NNSFC under Grants No. 12150610460 and by the supporting fund for foreign experts grant wgxz2022021L. AF was supported by RDF-22-02-079. LM was supported by an Australian Government Research Training Program (RTP) Scholarship and a Monash Graduate Excellence Scholarship (MGES). YW is supported by NNSFC under grant No. 12305112 and also by a starting grant from the Nanjing Normal University.
\end{acknowledgments}

\appendix

\section{Effective Potential}
\label{app:potential_rad_correc}
Following Refs.~\cite{Kobakhidze:2016mch,Kobakhidze:2017mru}, we construct a simple model that falls under the category of non-linearly realized electroweak symmetry.
The SM Higgs doublet belongs to the coset group $G_c = \textit{SU}(2)_L\times U(1)_Y/U(1)_{\text{EM}}$ and can be expressed as
\begin{equation}
    H(x) = \frac{r(x)}{\sqrt{2}}e^{i\theta^i (x)T^i}\begin{pmatrix}
    {0} \\
    {1}
    \end{pmatrix},
\end{equation}
where $i = 1-3$. The Higgs boson is a singlet field in the SM, denoted as $r(x)\sim (1, 1)_0$, and the fields $\theta^i (x)$ correspond to three would-be Goldstone bosons. The physical Higgs boson, $h$, is a fluctuation in $r$ around the vacuum expectation value of electroweak symmetry breaking, that is, $r = \langle r \rangle + h$, where $\langle r \rangle = v = 246\gev$.

The general tree-level Higgs potential for the SM singlet field, $r$, can be written as
\begin{equation}
    V_0 (r) = -\frac{\mu^2}{2} r^2 +\frac{\kappa}{3}r^3 +\frac{\lambda}{4}r^4.
\end{equation}
We add zero- and finite-temperature one-loop
Coleman-Weinberg
corrections
to the tree-level potential,
\begin{equation}
V(r, T) = V_0(r) + [V_{\text{CW}}(r) + V_T(r, T)]|_{m_i^2 \rightarrow m_i^2 + \Delta_T},
\end{equation}
and we replace all scalar and longitudinal gauge boson masses $m_i^2$ with the thermal masses $m_i^2 + \Delta_T$ (evaluated at lowest order), such that we are using the Parwani method~\cite{Parwani:1991gq} for Daisy resummation to address infrared divergences.\footnote{One could resum additional terms by matching to a three-dimensional effective field theory (see e.g.\ ref.~\cite{Ekstedt:2022bff}) but here we stick more closely to the procedure used in the original work on this idea.}
The formulas for $V_{\text{CW}}(r)$ and $V_T(r, T)$ and the thermal masses can be found in
the Appendix of
Ref.~\cite{Kobakhidze:2017mru}.  We apply Boltzmann suppression factors $e^{-m^2/T^2}$ to the Debye corrections as discussed in \cref{app:thermalParameters}. There are, of course, other possible models with supercooled FOPTs, including the classic Coleman-Weinberg model~\cite{Coleman:1973jx}.

The model parameters, namely $\mu$, $\kappa$, and $\lambda$, are constrained by the tadpole and on-shell mass conditions,
\begin{equation}
    \left.\frac{dV}{dr}\right|_{r=v} = 0,\quad\quad
    \left.\frac{d^2V}{dr^2}\right|_{r=v} = m^2_h,
\end{equation}
where $m_h \simeq 125\gev$. We use them to eliminate $\mu^2$ and $\lambda$ at the one-loop level through
\begin{subequations} \label{eq:one-loop-extraction}
\begin{align}
    \mu^2 & = \frac{1}{2} \left(m_h^2 + \kappa v + \frac{3}{v} \left. \dv{V_{\text{CW}}}{r} \right\vert_{v} - \left. \dv[2]{V_{\text{CW}}}{r} \right\vert_{v} \right) , \\
    \lambda & = \frac{1}{2 v^2} \left(m_h^2 - \kappa v + \frac{1}{v} \left. \dv{V_{\text{CW}}}{r} \right\vert_{v} - \left. \dv[2]{V_{\text{CW}}}{r} \right\vert_{v} \right) .
\end{align}
\end{subequations}
This requires an iterative procedure starting with the tree-level tadpole equations and repeatedly using one-loop extraction until convergence.
This leaves the cubic coupling $\kappa$ as the only free parameter.
The remaining cubic coupling $\kappa$ creates a barrier between minima
in the potential
and can lead to supercooling.
The requirement that the potential must be bounded from below ensures that $\lambda > 0$. On the other hand, by convention so that $\langle r \rangle > 0$, we choose $\kappa < 0$.

In addition to the particles stated in the Appendix of Ref.~\cite{Kobakhidze:2017mru}, we add radiative corrections from all remaining quarks and the muon and tau. The omitted states are always so light that we can treat them as radiation. We therefore account for $81$ effective degrees of freedom in the one-loop and finite-temperature corrections, leaving $25.75$ degrees of freedom from light particles that we treat as pure radiation.  Thus we add a final term to the effective potential:
\begin{equation}
    V_{\text{rad}}(T) = -\frac{\pi^2}{90} g'_* T^4 ,
\end{equation}
where $g'_* = 106.75 - 81 = 25.75$.

\section{Phase transition analysis} \label{app:phaseTransitionAnalysis}

We use \PT{}~\cite{Athron:2020sbe} to determine the phase structure (see ref.~\cite{Athron:2022jyi} for a discussion of uncertainties) and \TS{}~\cite{Athron:2023ts} to evaluate the phase history and extract the GW signal. The particle physics model considered in this study has at most one first-order phase transition, making phase history evaluation a simple matter of analyzing the single first-order phase transition. We use a modified version of \CT{}~\cite{Wainwright:2011kj} to calculate the bounce action during transition analysis.\footnote{The modifications are described in Appendix F of Ref.~\cite{Athron:2022mmm}. Most important are the fixes for underflow and overflow errors.}

\TS{} tracks the false vacuum fraction
\cite{Athron:2022mmm}
\begin{equation}
    P_f(T) = \exp\left[-\frac{4\pi}{3} v_w^3 \int_T^{T_c} \frac{\Gamma(T')dT'}{T'^4 H(T')} \left(\int_T^{T'} \! \frac{dT''}{H(T'')} \right)^{\!3} \right] \label{eq:Pf}
\end{equation}
as a function of temperature,\footnote{See Ref.~\cite{Athron:2022mmm} and section 4 of Ref.~\cite{Athron:2023xlk} for the assumptions implicit in \cref{eq:Pf}.}
where $v_w$ is the bubble wall velocity, $\Gamma$ is the bubble nucleation rate, $H$ is the Hubble parameter, and $T_c$ is the critical temperature at which the two phases have equal free-energy density.
This allows us to evaluate the GW power spectrum at the onset of percolation. Percolation occurs when the false vacuum fraction falls to $71\%$ \cite{doi:10.1063/1.1338506, LIN2018299, LI2020112815}. Thus we define the percolation temperature, $T_p$, through
\begin{equation}
    P_f(T_p) = 0.71 .
\end{equation}
This temperature will be used as the reference temperature for GW production. Additionally, we define the completion (or final) temperature, $T_f$, through
\begin{equation}
    P_f(T_f) = 0.01 , \label{eq:completion}
\end{equation}
as an indication of the end of the phase transition.

The quantities in \cref{eq:Pf} are estimated as follows. The bubble wall velocity $v_w$ is typically ultra-relativistic in the strongly supercooled scenarios we consider here, so we take $v_w \approx 1$. The bubble nucleation rate is estimated as \cite{Linde:1981zj}
\begin{equation}
    \Gamma(T) = T^4 \! \left(\frac{S(T)}{2\pi} \right)^{\!\!\frac32} \exp(-S(T)) ,
\end{equation}
where $S(T)$ is the bounce action. The Hubble parameter
\begin{equation}
    H(T) = \sqrt{\frac{8\pi G}{3} \rhotot(T)}
\end{equation}
depends on the total energy density \cite{Athron:2022mmm}
\begin{equation}
    \rhotot(T) = \rho_f(T) - \rhogs , \label{eq:rhotot}
\end{equation}
and $G = 6.7088 \! \times \! 10^{-39} \gev^{-2}$ is Newton's gravitational constant \cite{Wu:2019pbm}. The energy density of phase $\field_i$ is given by \cite{Espinosa:2010hh}
\begin{equation}
    \rho_i(T) = V(\field_i, T) - T \left. \pdv{V}{T} \right\vert_{\field_i(T)} ,
\end{equation}
where $V(\field_i, T)$ is the effective potential. The subscript $f$ in \cref{eq:rhotot} denotes the false vacuum and gs denotes the zero-temperature ground state of the potential.
Finally, the transition is analysed by evaluating the false vacuum fraction \cref{eq:Pf} starting near the critical temperature and decreasing the temperature until the transition completes. We define completion to be when $P_f(T_f) = 0.01$, and further check that the physical volume of the false vacuum is decreasing at $T_p$~\cite{Turner:1992tz, Athron:2022mmm}; that is,
\begin{equation}
    3 + T_p \left. \dv{\Vext}{T} \right\vert_{T_p} < 0 , \label{eq:Vphys}
\end{equation}
where $-\Vext$ is the exponent in \cref{eq:Pf}. This condition was empirically determined to be the strongest completion criterion of those considered in Ref.~\cite{Athron:2022mmm}, and continues to be in the models considered in this study.

\section{Thermal parameters} \label{app:thermalParameters}

The GW signal depends on several thermal parameters: the kinetic energy fraction $K$, the characteristic length scale $L_*$, the bubble wall velocity $v_w$, and a reference temperature $T_*$ for GW production. We take the reference temperature to be the percolation temperature $T_p$ because percolation necessitates bubble collisions \cite{Athron:2022mmm}. As explained above, we take $v_w \approx 1$ due to the strong supercooling. Specifically, we use the Chapman-Jouguet velocity \cite{Giese:2020rtr}
\begin{equation}
    v_w = v_\text{CJ} = \frac{1 + \sqrt{3\alpha (1 + c_{s,f}^2 (3\alpha - 1))}}{c_{s,f}^{-1} + 3 \alpha c_{s,f}} .
\end{equation}
The Chapman-Jouguet velocity is the lowest velocity for a detonation solution, and we expect more realistically that $v_w > v_\text{CJ}$ \cite{Laurent:2022jrs, Athron:2023xlk}. The choice $v_w = v_\text{CJ}$ is as arbitrary a choice as any fixed value of $v_w$, but has the benefit of always being a supersonic detonation. We note that a choice of $v_w < 1$ is required to estimate the kinetic energy fraction.

The kinetic energy fraction is the kinetic energy available to source GWs, divided by the total energy density $\rhotot$. We calculate this as \cite{Giese:2020znk}
\begin{equation}
    K = \frac{\pseudotrace_f(T_*) - \pseudotrace_t(T_*)}{\rhotot(T_*)} \kappa(\alpha, c_{s,f}, c_{s,t}) ,
\end{equation}
where
\begin{equation}
    \alpha = \frac{4(\pseudotrace_f(T_*) - \pseudotrace_t(T_*))}{3 w_f}
    \label{Eq:alpha}
\end{equation}
is the transition strength parameter, and the pseudotrace $\pseudotrace$ is given by \cite{Giese:2020rtr}
\begin{equation}
    \pseudotrace_i(T) = \frac14 \left(\rho_i(T) - \frac{p_i(T)}{c_{s,t}^2(T)} \right) .
\end{equation}
The pressure is $p = -V$, the enthalpy is $w = \rho + p$, and the speed of sound $c_s$ in phase $\field_i$ is given by
\begin{equation}
    c_{s,i}^2(T) = \left. \frac{\partial_T V}{T \partial_T^2 V} \right\vert_{\field_i(T)} . \label{eq:soundSpeed}
\end{equation}
This treatment of the kinetic energy fraction corresponds to model M2 of Refs.~\cite{Giese:2020rtr, Giese:2020znk}. We use the code snippet in the Appendix of Ref.~\cite{Giese:2020rtr} to calculate $\kappa(\alpha, c_{s,f}, c_{s,t})$; although $\kappa$ is independent of $c_{s,t}$ for a supersonic detonation. We note that $c_{s,f} \sim 1$ at very low temperature in our model if Boltzmann suppression is not employed. This is because the temperature-dependent contributions to the free energy density are dominated by the Debye corrections at low temperature. Hence, the free energy density in a phase scales roughly as $V(T) = aT^2+b$ at low temperature, where $a$ and $b$ are temperature independent. Consequently, the sound speed is roughly the speed of light by \cref{eq:soundSpeed}. However, applying Boltzmann suppression to the Debye corrections (as suggested in Ref.~\cite{Giese:2020znk}) corrects the sound speed back towards $c_s = 1/\sqrt{3}$ at low temperature. Specifically, for BP2 we find $c_{s,f}^2 \approx c_{s,t}^2 \simeq 1/3$ at $T_p = 0.1\gev$.

For turbulence, we take the efficiency coefficient $\kappa_{\text{turb}}$ to be 5\% and show it merely for comparison. Modeling the efficiency of the turbulence source is still an open research problem \cite{Athron:2023xlk}. While strong phase transitions could lead to significant rotational modes in the plasma \cite{Cutting:2019zws}, the resultant efficiency of GW production from turbulence is not yet clear.

We also consider a case where bubble collisions alone source GWs. In this case we ignore the sound wave and turbulence sources altogether and use $K = \alpha / (1 + \alpha)$ for the collision source. This assumes that the efficiency for generating GWs from the bubble collisions is maximal, which we take as a limiting case. We do not calculate the friction in the cubic potential so a proper estimate of the efficiency coefficient for the collision source is not possible.

We use the mean bubble separation $R_*$ for the characteristic length scale $L_*$. We calculate $R_*$ directly from the bubble number density, $n_B(T)$~\cite{Athron:2023xlk}: 
\begin{equation}
    R_*(T) = (n_B(T))^{-\frac13} = \left(T^3 \!\! \int_T^{T_c} \! dT' \frac{\Gamma(T') P_f(T')}{T'^4 H(T')} \right)^{\!\!-\frac13} .
\end{equation}
A common approach is to instead calculate
\begin{equation}
    \frac{\beta}{H} = T \dv{S}{T} ,
\end{equation}
which is a characteristic timescale for an exponential nucleation rate. The GW fits then implicitly map $\beta/H$ onto $R_*$ through
\begin{equation}
    R_* = (8\pi)^{\frac13} \frac{v_w}{\beta} . \label{eq:RbetaMap}
\end{equation}
However, an exponential nucleation rate is not appropriate for a strongly supercooled phase transition in the model we investigate. Further, $\beta/H$ becomes negative below $T_\Gamma$ (i.e.~below the minimum of the bounce action). While alternative mappings exist for a Gaussian nucleation rate \cite{Megevand:2016lpr, Ellis:2018mja}, usually \cref{eq:RbetaMap} is inverted in GW fits when using $\beta/H$.

We also incorporate the suppression factor
\begin{equation}
    \Upsilon(\tausw) = 1 - \frac{1}{\sqrt{1 + 2 H_*\tausw}} ,
\end{equation}
in our GW predictions, which arises from the finite lifetime of the sound wave source \cite{Guo:2020grp}. We use the shorthand notation $H_* = H(T_*)$. The timescale $\tausw$ is estimated by the shock formation time $\tausw \sim L_*/\Uf$, where $\Uf = \sqrt{K \rho_f / \overline{w}}$ and $\overline{w}$ is the average enthalpy density \cite{Athron:2023xlk}.

\section{Redshifting} \label{app:redshift}

The GW spectrum we see today is redshifted from the time of production. The frequency and amplitude scale differently (see refs.~\cite{Cai:2017tmh, Athron:2023xlk}). The redshift factors are obtained using conservation of entropy and the assumption of radiation domination. Here, we avoid the latter assumption, thus our redshift factors may look unfamiliar.

Frequencies redshift according to \cite{Athron:2023xlk}
\begin{equation}
    f_0 = \frac{a_1}{a_0} f_1 = \mathcal{R}_f f_1 , \label{eq:redshiftFreq-fundamental}
\end{equation}
where $a$ is the scale factor of the Universe, and we have defined the redshift factor for frequency, $\mathcal{R}_f$. Using conservation of entropy,
\begin{equation}
    a_0^3 s_0 = a_1^3 s_1 ,
\end{equation}
where $s$ is the entropy density, $\mathcal{R}_f$ becomes
\begin{equation}
    \mathcal{R}_f = \left(\frac{s_0}{s_1} \right)^{\!\frac13} .
\end{equation}
The number of entropic degrees of freedom at the current temperature $T_0 = 2.725 \unit{K} = 2.348\tentothe{-13}\gev$ \cite{Fixsen:2009ug} is
\begin{equation}
    g_s(T_0) = 2 + \frac{7}{11} N_\text{eff} ,
\end{equation}
where $N_\text{eff} = 3.046$ is the effective number of neutrinos. The entropy density today is $s_0 = 2.237\tentothe{-38}\gev^3$ which we computed from the temperature derivative of the effective potential. Because the frequency $f_1$ is typically determined using quantities expressed in GeV, we apply the unit conversion $1\gev = 1.519\tentothe{24} \unit{Hz}$ to express the dimensionful redshift factor for frequency as
\begin{equation}
    \mathcal{R}_f = 4.280\tentothe{11} \frac{\!\unit{Hz}}{\!\unit{GeV}} \left(\frac{1\gev^3}{s_1} \right)^{\!\frac13} . \label{eq:redshiftFreq}
\end{equation}

The amplitude redshifts according to \cite{Athron:2023xlk}
\begin{equation}
    \Omega_0 h^2 = \left(\frac{a_1}{a_0} \right)^4 \left(\frac{H_1}{H_0} \right)^2 \Omega_1 h^2 = \mathcal{R}_\Omega \Omega_1 ,
\end{equation}
where $H$ is the Hubble parameter, and we have defined the redshift factor for the amplitude, $\mathcal{R}_\Omega$, which absorbs the factor $h^2$. The Hubble parameter today is $H_0 = 100 h \unit{km} \unit{s}^{-1} \unit{Mpc}^{-1}$. Using $h = 0.674 \pm 0.005$ \cite{ParticleDataGroup:2020ssz} and again converting from Hz to GeV, we have $H_0 = 1.438\tentothe{-38}\gev$. Thus, the dimensionless redshift factor for amplitude is
\begin{equation}
    \mathcal{R}_\Omega = 1.384\tentothe{33} \left(\frac{\unit{GeV}^3}{s_1} \right)^{\!\frac43} \left(\frac{H_1}{\!\unit{GeV}} \right)^{\!2} .
\end{equation}

If the GWs are produced at temperature $T_*$ and reheating increases the temperature to $\Treh$ in the true vacuum, we take $s_1 = s_t(\Treh)$ and $H_1 = H(T_*)$. We assume conservation of entropy in the true vacuum, where $s_t = -\partial V / \partial T|_{\field_t}$, such that adiabatic cooling occurs for the temperature range $T_1 = \Treh$ to $T_0$. We use $T_*$ in the Hubble parameter because $\rho_f(T_*) = \rho_t(\Treh)$ by the definition of $\Treh$. We find that the redshift factors $\mathcal{R}_f$ and $\mathcal{R}_\Omega$ are within 1\% of the values obtained when assuming radiation domination, at least for BP1 and BP2. This demonstrates that radiation domination is a good assumption in the true vacuum in this model.

\section{Gravitational waves}\label{app:GWsFits}

We consider three contributions to the GW signal: bubble collisions, sound waves in the plasma, and magnetohydronamic turbulence in the plasma. For simplicity, we consider two scenarios: 1) non-runaway bubbles, where GWs are sourced purely by the plasma because the energy stored in bubble walls is dissipated into the plasma; and 2) runaway bubbles, where GWs are sourced purely by the energy stored in the bubble walls. We do not consider the fluid shells in this latter case. In the following, we reverse common mappings such as $R_* = (8\pi)^{\frac13} v_w/\beta$ and $K = \kappa \alpha / (1 + \alpha)$ to generalise the GW fits beyond assumptions made in the original papers. This generalisation comes at the cost of further extrapolation, beyond what is already inherent in using such fits. We also use our redshift factors derived in \cref{app:redshift} instead of the radiation domination estimates. We refer the reader to the reviews in Refs.~\cite{Caprini:2015zlo, Caprini:2019egz, Athron:2023xlk} for further discussions on the GW fits listed below.

We use the recent GW fit for the collision source from Ref.~\cite{Lewicki:2022pdb}. The redshifted peak amplitude is
\begin{align}
    \Omega_{\mathrm{coll}}(f) & = \mathcal{R}_\Omega A \left(\frac{H_* R_*}{(8\pi)^{\frac13} v_w}\right)^{\!2} \! K^2 \,S_{\mathrm{coll}}(f) , \label{eq:GWamp_coll}
\end{align}
and the spectral shape is
\begin{equation}
    S_{\mathrm{coll}}(f)=\frac{(a+b)^{c}}{\left[b\left(\frac{f}{f_\text{coll}}\right)^{\!-\frac{a}{c}}+a\left(\frac{f}{f_\text{coll}}\right)^{\!\frac{b}{c}}\right]^{c}} .
\end{equation}
The redshifted peak frequency is
\begin{equation} \label{eq:fpeak-coll}
    f_\text{coll} = \mathcal{R}_f \left(\frac{0.77 (8\pi)^{\frac13} v_w}{2\pi R_*} \right) .
\end{equation}
The fit parameters $A$, $a$, $b$, $c$, and $f_p$ (the peak frequency before redshifting) can be found in Table I in Ref.~\cite{Lewicki:2022pdb}; specifically the $T_{rr} \propto R^{-3}$ column for bubbles. We normalised the spectral shape by moving $A$ into \cref{eq:GWamp_coll} as an explicit factor. We have mapped $f_p/\beta$ onto $1/R_*$ in \cref{eq:fpeak-coll}.

For the sound wave source, we use the GW fits in the sound shell model \cite{Hindmarsh:2016lnk} from Refs.~\cite{Hindmarsh:2019phv, Gowling:2022pzb}. The redshifted peak amplitude is
\begin{equation} \label{eq:GWamp_sw}
    h^{2} \Omega_{\mathrm{sw}}(f) = 3 \mathcal{R}_\Omega K^2 \left(\frac{H_* R_*}{c_{s,f}}\right) \frac{M\left(s, r_b, b\right)}{\mu_f(r_b)} \Upsilon(\tausw) \tilde{\Omega}_{\mathrm{gw}} ,
\end{equation}
with spectral shape
\begin{equation}
    M\left(s, r_{\mathrm{b}}, b\right)=s^{9}\left(\frac{1+r_b^{4}}{r_b^4+s^4}\right)^{\!\frac{9-b}{4}}\left(\frac{b+4}{b+4-m+m s^2}\right)^{\!\frac{b+4}{2}} ,
\end{equation}
where
\begin{align}
    &m =\left(9 r_b^{4}+b\right) /\left(r_b^{4}+1\right) ,\\
    &s = f/f_p,\\
    &r_b = f_b/f_p, \\
    &\mu_f(r_b) =4.78-6.27 r_b+3.34 r_b^2 .
\end{align}
In \cref{eq:GWamp_sw} we have used $\tau_c \sim R_* / c_{s,f}$ for the autocorrelation timescale \cite{Caprini:2019egz}, hence the factor $1/c_{s,f}$.
We take $\tilde{\Omega}_{\mathrm{gw}} = 0.01$ in accordance with Table IV of Ref.~\cite{Hindmarsh:2017gnf}, and $b = 1$. The breaks in the power laws are governed by the mean bubble separation and the fluid shell thickness, which respectively correspond to the redshifted frequencies \cite{Hindmarsh:2017gnf}
\begin{equation}
    f_b = 1.58 \mathcal{R}_f \left(\frac{1}{R_*} \right) \left(\frac{z_p}{10} \right) \label{eq:fpeak-bubsep}
\end{equation}
and
\begin{equation}
    f_p = 1.58 \mathcal{R}_f \left(\frac{1}{R_* \Delta_w} \right) \left(\frac{z_p}{10} \right) .
\end{equation}
The length scale for the fluid shell thickness is roughly \cite{Hindmarsh:2019phv}
\begin{equation}
    R_* \Delta_{w} \approx R_* \abs{v_{w}-c_{s, f}} / v_{w} ,
\end{equation}
although see Ref.~\cite{Athron:2023xlk} for further discussion.
We take the dimensionless wavenumber at the peak to be $z_p = 10$ which is applicable for the supersonic detonations we consider~\cite{Hindmarsh:2017gnf, Hindmarsh:2019phv}. We note that a more recent analysis --- taking into account the scalar-driven propagation of uncollided sound shells --- reproduces the causal $f^3$ scaling below the peak of the GW signal found in numerical simulations \cite{Cai:2023guc}. Additionally, the spectral shape should depend on the thermal parameters \cite{Cai:2023guc, Ghosh:2023aum}.

Finally, for the turbulence fit, we use the fit from Ref.~\cite{Caprini:2010xv} based on the analysis in Ref.~\cite{Caprini:2009yp}, using $L_* = R_*$ rather than $L_* \sim 2 v_w/\beta$. The redshifted peak amplitude is
\begin{equation}
    \Omega_\mathrm{turb}(f) = 9.0 \mathcal{R}_\Omega \left(H_{*}R_*\right) (\kappa_{\text{turb}} K)^{\frac{3}{2}} S_{\!\mathrm{turb}}(f) ,
\end{equation}
with unnormalised spectral shape
\begin{equation}
    S_{\mathrm{turb}}(f)=\frac{\left(f / f_{\mathrm{turb}}\right)^{3}}{\left(1+f / f_{\mathrm{turb}}\right)^{\frac{11}{3}}\left(1 + 8\pi f / (\mathcal{R}_f H_*) \right)} .
\end{equation}
We take $\kappa_{\mathrm{turb}} = 0.05$ as a conservative estimate of the turbulence source in a strongly supercooled transition.
The redshifted peak frequency is \cite{Caprini:2009yp}
\begin{equation}
    f_\text{turb} = \mathcal{R}_f \frac{3.5}{R_*} . \label{eq:fpeak-turb}
\end{equation}
Note that we have not assumed $f_\text{turb} \sim \pi\beta/(2v_w)$ like was done in Ref.~\cite{Caprini:2010xv}.

\bibliography{refs}
\bibliographystyle{JHEP}
\end{document}